\documentclass[conference,10pt]{IEEEtran}

\usepackage{amssymb}
\usepackage{multirow}
\usepackage{amsmath}
\usepackage{amssymb}
\usepackage{graphicx}
\usepackage{enumerate}
\title{System-Level Performance of mmWave Cellular Networks for Urban Micro Environments}

\author{
\IEEEauthorblockA{Nadisanka~Rupasinghe\IEEEauthorrefmark{1}, Yuichi Kakishima\IEEEauthorrefmark{2}, \. {I}smail~G\"uven\c{c}\IEEEauthorrefmark{1}
\IEEEauthorblockA{\IEEEauthorrefmark{1}Dept. of Electrical and Computer Engineering, North Carolina State University, Raleigh, NC}
\IEEEauthorblockA{\IEEEauthorrefmark{2}DOCOMO Innovations Inc., Palo Alto, CA}
Email: {\tt rprupasi@ncsu.edu, kakishima@docomoinnovations.com, iguvenc@ncsu.edu}}%
}

\begin{document}
\maketitle
\begin{abstract}
   In this paper, we investigate the propagation coupling loss (captures all sources of attenuation between serving cell and mobile station (MS)) and geometry metric (GM) (downlink average
signal-to-interference plus noise ratio) performance of mmWave cellular networks for outdoor and indoor MSs, considering urban micro (UMi) environments. Based on these studies, we identify effective mmWave frequency bands for cellular communication. We consider 3GPP compliant system-level simulations with two power allocation schemes: 1) transmit power scaled with communication
bandwidth, and 2) constant total transmit power. Simulation results show that with scaled transmit power allocation, GM performance degradation is small: $20\%$ of MSs experience GM less than $0$~dB at all mmWave frequencies considered, for outdoor MSs. With constant Tx power allocation, $20\%$ of MSs experience GM less than $0$ dB for frequencies up to $30$ GHz. Furthermore, $35\%$ ($48\%$) of outdoor MSs experience GM performance less than $0$ dB at $60$ GHz ($100$ GHz). On the other hand, for indoor MSs, even with scaled Tx power allocation, favorable GM performance is
observed only at low frequencies, i.e., $2$~GHz.

\end{abstract}
\vspace{-3 mm}
\section{Introduction}
{\let\thefootnote\relax\footnotetext{To appear in URSI GASS 2017 conference.}}Achieving higher system capacity and higher data rates are two major goals in fifth-generation (5G) mobile communication systems. To this end, extending the operation of 5G systems to millimeter-wave (mmWave) bands is critical due to the availability of large amount of bandwidth. However, before extending cellular communication to mmWave bands, it is important to develop accurate and flexible mmWave propagation models, and investigate achievable performance with mmWave transmission by using those models.

There are several recent efforts to develop mmWave propagation models for 5G networks \cite{mmMAGIC, NIST, CIwhite, RP151606}. A mmWave channel model is developed in \cite{Rapport1}, based on extensive channel measurements in $28$ GHz, $38$ GHz, $60$ GHz, and $73$ GHz mmWave bands, considering peer-to-peer and vehicular scenarios. A measurement based \cite{6387266} path loss (PL) model is presented in \cite{6785327}, along with a distance dependent line-of-sight (LoS) probability model. Another measurement based mmWave PL model is introduced in \cite{6363950} considering $54$-$59$ GHz and $61$-$66$ GHz mmWave bands. In that, ultra wide band pseudo noise signal with bandwidth of $1.2$ GHz is used for channel sounding. Recently, in collaboration with fourteen different institutions, three mmWave PL models are developed in \cite{CIwhite}: 1) \emph{close-in} (CI) free space reference distance model, 2) \emph{alpha-beta-gamma} (ABG) model, and 3) CI free space reference distance model with frequency dependent PL exponent (CIF), based on channel measurements and ray tracing data.

Even though there are several efforts in developing mmWave propagation models, there are limited investigations on system-level propagation performance of mmWave cellular systems. Although it is generally understood that mmWave frequency bands tend to achieve poor propagation performance e.g. due to increased PL, oxygen absorption, it is not quantitatively understood which mmWave bands are effectively available for cellular network operation.

 In this paper, we quantitatively analyze mmWave propagation characteristics in multi-cell environments using 3GPP based system-level simulations. In particular, we focus on propagation coupling loss ($\rm {CL}$) and geometry metric (GM) performance of outdoor and indoor mobile stations (MSs) in urban micro (UMi) environments. The propagation $\rm {CL}$ provides information regarding the attenuation the desired signal undergoes when traveling from serving cell to MS. On the other hand, the  GM performance captures the quality of the desired signal at the MS by taking into consideration the  interference from other transmissions and \emph{additive white gaussian noise} (AWGN). We consider mmWave propagation models proposed in \cite{CIwhite} for our analysis. These models are also the baseline propagation models for 3GPP mmWave channel models. Further, we consider two power allocation schemes in our investigation to understand the possibility of compensating for losses incurred at mmWave frequencies. The rest of the paper is organized as follows. In Section~\ref{sec:PLmodels}, we review mmWave propagation models considered in our analysis. System-level simulation results for propagation CL and GM performance are discussed in Section~\ref{sec:PLEval}. Finally, Section~\ref{sec:Conclusion} provides concluding remarks.
%Finally, from our analysis, we identify effective mmWave frequencies for cellular operation.
\section{mmWave Propagation Models} \label{sec:PLmodels}

The mmWave propagation models recently developed in \cite{CIwhite} based on extensive channel measurements and ray tracing simulations are considered for our investigation. In this section, we review those propagation models to obtain a better insight about how specific propagation behaviors in mmWave frequencies are captured within those models.
\subsection{Path Loss for LoS and NLoS}
All PL models are functions of transmission frequency and distance, and are applicable for $0.5 - 100$ GHz frequency range. For our analysis, we consider CI model as LoS PL model and ABG model as non-LoS (NLoS) PL model. The LoS PL model from CI model is defined as \cite{CIwhite}:
\begin{align} \label{equ:CI}
\small
PL^{\rm {LoS}}(f_{\rm c},d) = FSPL(f_{\rm c}) + 21 \log_{10}(d) +  \mathcal{X}_{\sigma}^{\rm {LoS}},
\end{align} \normalsize
where free space path loss (FSPL) is defined as
\begin{align*}
FSPL(f_{\rm c}) = 20\log_{10}\left(\frac{4\pi f_{\rm c}}{c}\right),
\end{align*} and $d$, $f_{\rm c}$ and $c$ are the distance between transmitter and receiver (in m), operating frequency (in Hz), and the speed of light (in m/s), respectively. Shadow fading for the CI model ($\mathcal{X}_{\sigma}^{\rm{LoS}}$) is normally distributed with $\mathcal{N} (0, 3.76^{\rm{2}})$. The NLoS PL model from ABG model is given by
\vspace{-2.5 mm}

 \small \begin{align} \label{equ:ABG}
PL^{\rm{NLoS}}(f_{\rm{c}},d) & = 10\alpha \log_{10}(d) + \beta + 10\gamma \log_{10}(f_{\rm{c}})+\mathcal{X}_{\sigma}^{\rm{NLoS}}.
\end{align} \normalsize
Here, $\alpha=3.53$ captures how the path loss varies with distance, $\beta=22.4$ is a floating offset value in dB, and $\gamma=2.13$ captures path loss variation with frequency (in GHz). Shadow fading is normally distributed and captured by the term $\mathcal{X}_{\sigma}^{\rm{NLoS}}$ $\sim \mathcal{N} (0, 7.82^{\rm{2}})$. For calculating LoS probability, \cite{CIwhite} proposes to consider 3GPP LoS probability model in \cite{3GPPRP36873}.

\subsection{Outdoor-to-Indoor Penetration Loss}
The Outdoor-to-Indoor (O2I) penetration loss ($L_{\rm O2I}(f_{\rm c})$) considered in our analysis can be given as \cite{CIwhite}
\begin{align} \label{equ:Penetration}
L_{\rm O2I}(f_{\rm c}) = L_{\rm tw}(f_{\rm{c}}) + L_{\rm in} + \mathcal{X}_{\sigma}^{\rm {O2I}},
\end{align} where $L_{\rm tw}(f_{\rm{c}})$, and $L_{\rm in}$ are building penetration loss, and loss due to signal traveling inside the building, respectively. A normally distributed random loss, $\mathcal{X}_{\sigma}^{\rm {O2I}}$ $\sim \mathcal{N} (0, \sigma_{\rm O2I}^{\rm{2}})$ is also introduced. Note here that the penetration loss through the external wall ($L_{\rm tw}(f_{\rm{c}})$) depends on $f_{\rm{c}}$ and in \cite{CIwhite}, models for this frequency dependent penetration loss are provided for standard multi-pane glass ($L_{\rm g}$), infrared reflective (IRR) glass ($L_{\rm IRRg}$) and concrete ($L_{\rm c}$) materials. Then the composite penetration loss is obtained by considering a weighted average of the transmission through two different materials mentioned previously. Two variants of the composite loss model are proposed \cite{CIwhite}; 1) low loss model ($L^{\rm Low}_{\rm tw}$), and 2) high loss model ($L^{\rm High}_{\rm tw}$). The penetration loss through wall for each model is then defined as,
\vspace{-4.5mm}

\small \begin{align*}
L^{\rm Low}_{\rm tw} = 5-10\log_{10}\left(0.3\times 10^{\frac{-L_{\rm g}}{10}} + 0.7\times 10^{\frac{-L_{\rm c}}{10}}\right) + \mathcal{X}_{\sigma, L}^{\rm {O2I}} 
\end{align*} \begin{align*}
L^{\rm High}_{\rm tw} = 5-10\log_{10}\left(0.7\times 10^{\frac{-L_{\rm IRRg}}{10}} + 0.3\times 10^{\frac{-L_{\rm c}}{10}}\right) + \mathcal{X}_{\sigma, H}^{\rm {O2I}}.
\end{align*} 

\normalsize

\vspace{-2.5mm}
For $L^{\rm Low}_{\rm tw}$, and $L^{\rm High}_{\rm tw}$, $\sigma_{\rm O2I}^2$ is defined as $3$, and $5$ \cite{CIwhite}, respectively. Then, $L_{\rm tw}(f_{\rm{c}})+ \mathcal{X}_{\sigma}^{\rm {O2I}}$ is calculated as,
\vspace{-2.5mm}

\small \begin{align}
 L_{\rm tw}(f_{\rm{c}})+ \mathcal{X}_{\sigma}^{\rm {O2I}} = 10\log_{10}\left(0.5\times 10^{\frac{L^{\rm Low}_{\rm tw}}{10}} + 0.5\times 10^{\frac{L^{\rm High}_{\rm tw}}{10}}\right).
\end{align}
%\vspace{-2.5mm}

\normalsize

\subsection{Oxygen Absorption Loss}
We consider a frequency dependent oxygen absorption loss model for our analysis. The oxygen absorption loss, $L_{\rm OA}(f_{\rm{c}},d)$ can be given as \begin{align} \label{equ:OA}
L_{\rm OA}(f_{\rm{c}},d) = \delta (f_{\rm c}) \times d,
\end{align} where $\delta (f_{\rm c})$ is a frequency dependent loss factor.

\section{ System-Level Performance Analysis Using 3GPP based System-Level Simulations} \label{sec:PLEval}
\begin{table}
\small
\centering
\caption{System-level simulator configuration.}
\begin{tabular}{ | c | c | }
  \hline			
  \textbf{Parameter} 			&  \textbf{Value} \\ \hline
  Deployment scenario       & $19$ BSs, 3 sectors / BS \\ \hline
  ISD            			&  $200$ m (3D-UMi)\\ \hline       			
  BS antenna height ($h_{\rm BS}$)     			& $10$ m (3D-UMi) \\ \hline
  MS distribution 	   & Outdoor only and indoor only  \\ \hline
  Avg. no. UEs per sector       & $10$ \\ \hline
  Noise level ($\rm N_0$)        	& $-174$ dBm/Hz \\ \hline
  Noise figure        	& $9$ dB \\ \hline
\end{tabular}
\label{tab:config}
\vspace{1.5mm}
\end{table}

In this section, we quantitatively investigate geometry and propagation CL performance in a mmWave cellular network using a 3GPP compliant system-level simulator. We consider 3-tier cell layout with $19$ base stations (BSs) each with $3$ sectors (all together $57$ sectors). A wrap-around architecture is considered to have similar interference impact in all the cells. UEs are dropped uniformly and randomly within the given area.  The BS is equipped with a uniform linear antenna array (ULA) having $10$ antenna elements and generates a vertical beam with a $10.2$ degree half power beamwidth, and $17.6$ dB maximum gain. The beam is electrically down tilted by $102$ degrees (elevation angle from zenith) for transmission. MS consists of a single antenna element. Parameter configurations considered are summarized in Table~\ref{tab:config} (see also Table 8.2-2 in \cite{3GPPRP36873}). Further, as shown in Table~\ref{power}, we consider different bandwidths (BW) for different $f_{\rm{c}}$ values. This is an important consideration since one of the main motivations to move to mmWave spectrum is the availability of large bandwidth. Oxygen absorption loss, $L_{\rm OA}(f_{\rm{c}},d)$ is significant only at $60$~GHz with $\delta=15$~dB/km \cite{CIwhite} and negligible for all the other $f_{\rm c}$ considered in the analysis. The system-level simulator is calibrated with 3GPP requirements as described in \cite{3GPPRP36873}.

We consider two power allocation schemes: 1) transmit power ($P_{\rm {Tx}}$) is available proportionally with the BW, and 2) constant $P_{\rm {Tx}}$ irrespective of the available BW. Table~\ref{power} summarizes $P_{\rm {Tx}}$ for different~$f_{\rm c}$. Finally, for all our investigations, we consider UMi environment corresponding to the street canyon environment.

\begin{table}
\small
\centering
\caption{Bandwidth and Power allocation.}
\begin{tabular}{|c|c|c|c|c|c|}
\hline
 {$\boldsymbol{f_{\rm{c}}}$ (\textbf{GHz})}      & \textbf{2}  & \textbf{10}  & \textbf{30} & \textbf{60}  & \textbf{100} \\ \hline
 {\textbf{BW (MHz)}} & 20 & 300  & 500 & 1000 & 2000  \\ \hline
 \multirow{4}{*} {$P_{\rm {Tx}}$} &\multicolumn {5}{|c|}{\textbf{Scaled Tx power}}\\ \cline{2-6}
 & 44.0 & 55.8 & 58.0 &  61.0 & 64.0 \\ \cline{2-6}
%  %&{\textbf{UMa}}  & 49.0 & 60.8 & 63.0 &  66.0 & 69.0 \\ \cline{2-7}
 &\multicolumn {5}{|c|}{\textbf{Constant Tx power}}\\ \cline{2-6}
 (dBm)& \multicolumn {5}{|c|}{44.0}\\ \hline
  %&{\textbf{UMa}}  & 49.0 & 49.0  & 49.0 &  49.0 & 49.0 \\ \hline
\end{tabular}
\label{power}
%\vspace{-4mm}
\end{table}

\subsection{Coupling Loss Performance Analysis } \label{sec:UMiCL}

In this section, we analyze propagation $\rm {CL}$ performance for outdoor and indoor MSs. CL captures all sources of attenuation, i.e., due to propagation, antenna radiation, of the signal between serving cell and MS \cite{LTEBook}. Since $\rm {CL}$ does not depend on $P_{\rm {Tx}}$, $\rm {CL}$ is the same for both Tx power allocation schemes considered here. $\rm {CL}$ is defined as:\begin{align} \label{equ:CLEq}
{\rm {CL}}  = G_{\rm{Tx}}& + G_{\rm{Rx}} \\
&- \underbrace{\left(PL(f_{\rm{c}},d)+ L_{\rm O2I}(f_{\rm{c}}) + L_{\rm OA}(f_{\rm{c}},d) - G_{\rm{sm}} \right)}_{\rm Link \ Loss}, \notag \end{align} \normalsize where $G_{\rm{Tx}}$ and $G_{\rm{Rx}}$ are transmit and receive antenna gains, respectively. The gain due to multipath transmission is captured in $G_{\rm{sm}}$, which  is affected by the antenna array geometry \cite{3GPPRP36873}. In order to analyze $\rm {CL}$, we define link loss as,

\vspace{-6mm}
\small
\begin{align} \label{equ:linkloss}
{\rm {Link~Loss}} = PL(f_{\rm{c}},d) + L_{\rm O2I}(f_{\rm{c}}) + L_{\rm OA}(f_{\rm{c}},d) - G_{\rm{sm}}.
\end{align} \normalsize With increasing $PL(f_{\rm{c}},d)$, $L_{\rm O2I}(f_{\rm{c}})$, and $L_{\rm OA}(f_{\rm{c}},d)$, link loss in (\ref{equ:linkloss}) increases. Hence, $\rm {CL}$ in (\ref{equ:CLEq}) decreases with increasing link loss (larger negative value). In the subsequent sections, we study how each of these factors affect $\rm {CL}$ performance.

\subsubsection{Coupling Loss for Outdoor MSs} \label{outCL}
\begin{figure}
%\vspace{-3mm}
\begin{center}
\includegraphics[width=3.0in]{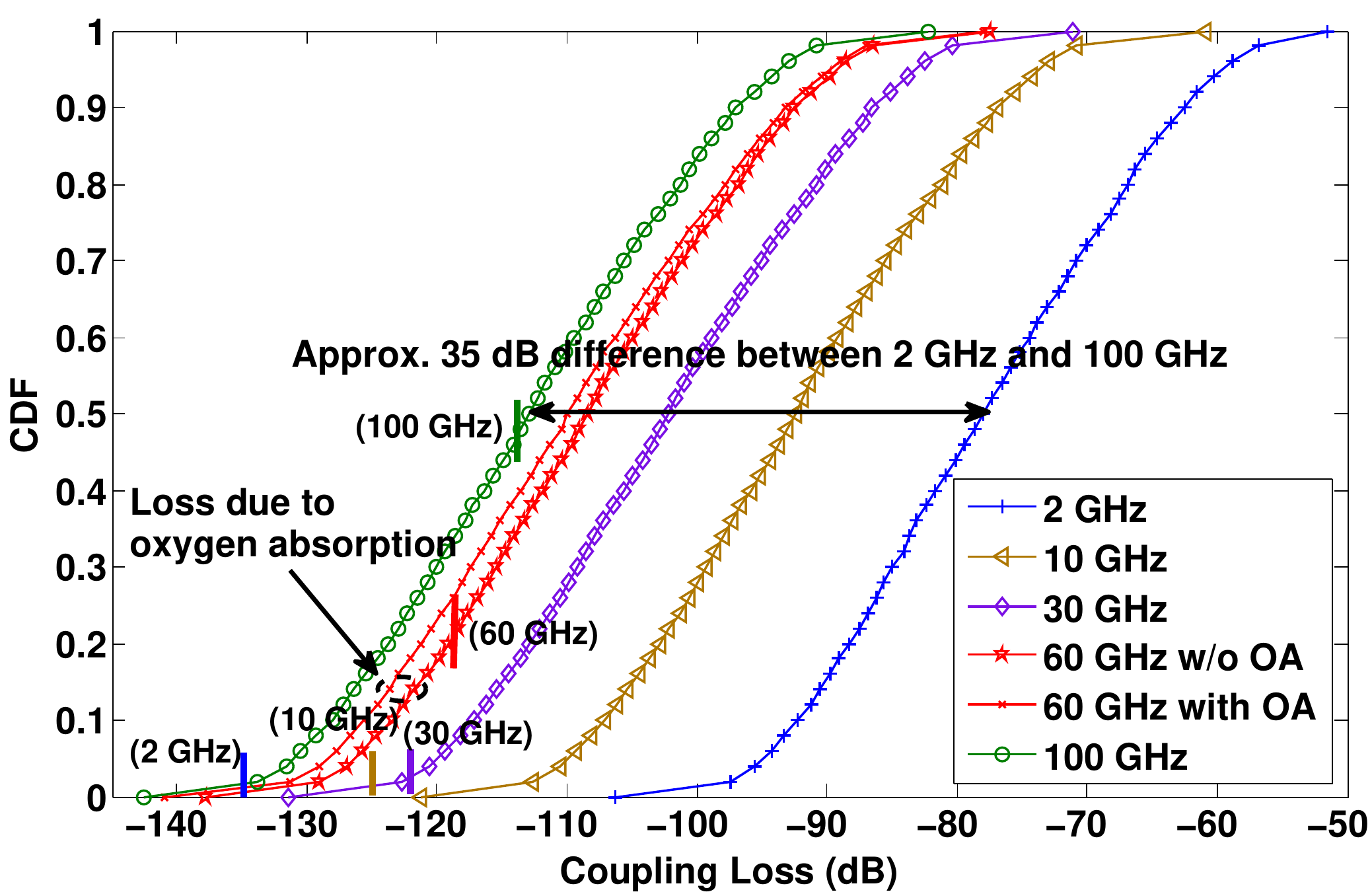}
\end{center}
%\vspace{-1mm}
\caption{ CDFs of coupling loss for outdoor MSs.}
\label{fig:OutCL}
%\vspace{-3mm}
\end{figure}

In Fig.~\ref{fig:OutCL}, cumulative distribution functions (CDFs) of $\rm {CL}$ for outdoor MSs with different $f_{\rm{c}}$ values are presented. As can be observed from that, $\rm {CL}$ decreases (link loss increases) with $f_{\rm{c}}$. This is because of the high PL experienced at higher frequencies. For example, approximately $35$ dB difference in $\rm {CL}$ can be seen between $2$ GHz and $100$ GHz at $0.5$ CDF point. Further, an additional loss of about $3$~dB at $60$~GHz (compared to w/o $L_{\rm OA}(f_{\rm{c}},d)$) can be observed due to oxygen absorption.

Fig.~\ref{fig:OutCL} also shows $\rm {CL}$ values for which received signal-to-noise power ratio (SNR) is zero (${\rm {CL}_{SNR=0}} = -P_{\rm Tx}+N_0$, where $N_0$ is AWGN) using colored line segments with colors corresponding to $f_{\rm c}$, for constant total Tx power allocation. For a particular $f_{\rm{c}}$, when $\rm {CL}$ is less than the ${\rm {CL}_{SNR=0}}$, the system is operating in \emph{noise-limited} condition (received signal power is below constant noise level, $N_0$). On the other hand, when observed $\rm {CL}$ is greater than the ${\rm {CL}_{SNR=0}}$ (received signal power is larger than $N_0$), the system tends to operate in \emph{interference-limited} condition. As per Fig.~\ref{fig:OutCL}, for outdoor MSs, up to $30$~GHz, system sensitivity to noise is not significant (\emph{interference-limited}) whereas at $60$~GHz, and $100$~GHz, system noise sensitivity increases.

%\vspace{-1mm}

\subsubsection{Coupling Loss for Indoor MSs}\label{inCL}
\begin{figure}
%\vspace{-3mm}
\begin{center}
\includegraphics[width=3.0in]{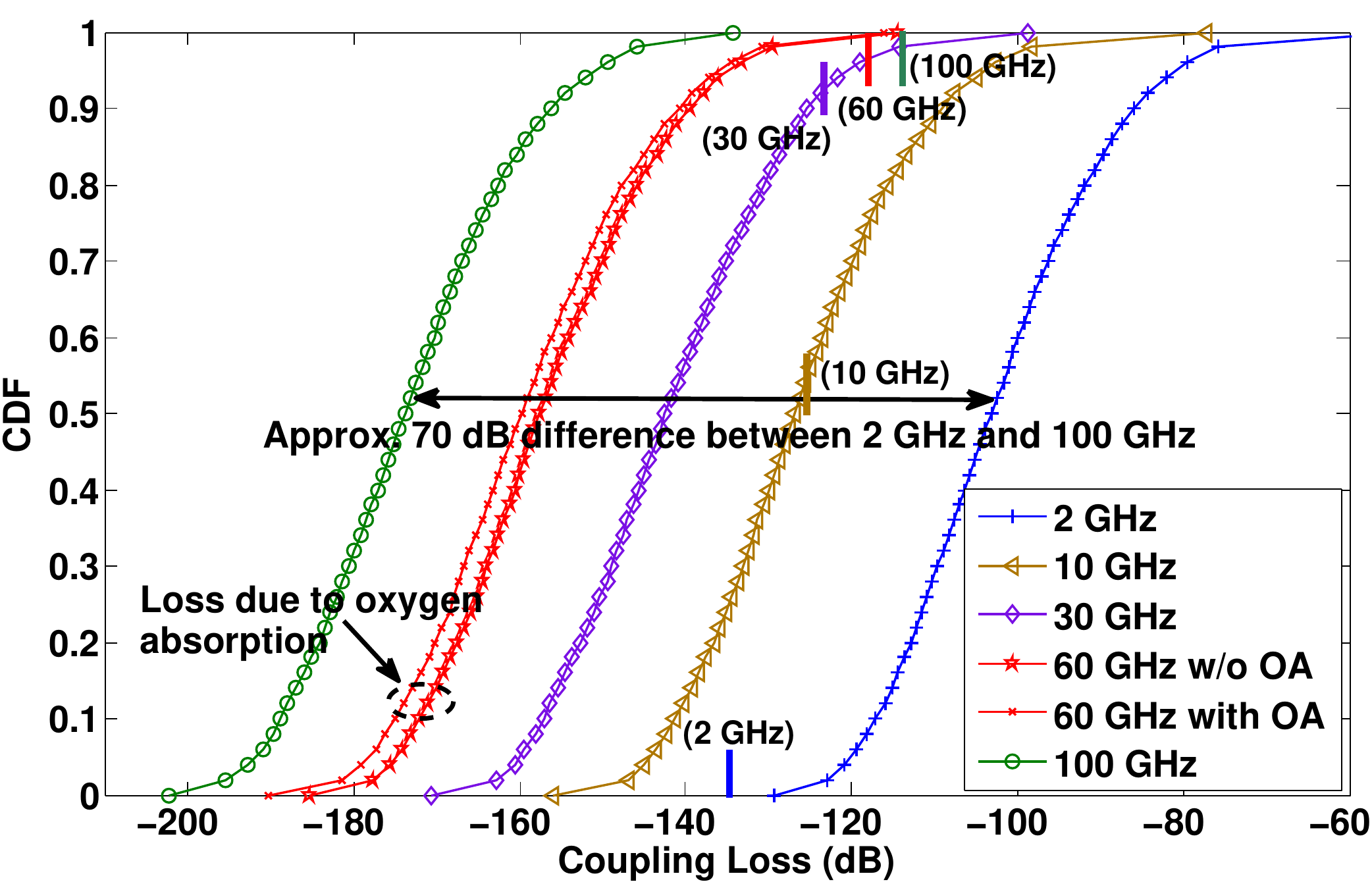}
\end{center}
%\vspace{-1mm}
\caption{CDFs of coupling loss for indoor MSs.}
\label{fig:InCL}
%\vspace{-1mm}
\end{figure}

$\rm {CL}$ distribution for indoor MSs is presented in Fig.~\ref{fig:InCL}. As can be observed, there is a $\rm {CL}$ difference of about $70$~dB between $2$ GHz and $100$ GHz at $0.5$ CDF point. Compared to outdoor MSs,  $\rm {CL}$ performance for indoor MSs is degraded due to the additional $L_{\rm O2I}(f_{\rm{c}})$.
%For example, $f_{\rm{c}} = 100$ GHz is subject to $35$ dB higher loss compared to $f_{\rm{c}} = 2$ GHz.

Further, it can clearly be seen from Fig.~\ref{fig:InCL}, at $60$~GHz and $100$~GHz, almost all the indoor MSs are in \emph{noise-limited} condition (colored line segments with colors corresponding to $f_{\rm c}$) due to $L_{\rm O2I}(f_{\rm{c}})$ experienced by indoor MSs. In addition, indoor MSs also suffer additional loss of about $4$~dB at $60$~GHz due to oxygen absorption which is slightly higher than that experienced by outdoor MSs. This is because, signal has to travel some additional distance inside the building also ($L_{\rm OA}(f_{\rm{c}},d)$ depends on distance traveled).

\subsection{Geometry Metric Performance Analysis} \label{sec:UMiGeo}
In this section we evaluate GM performance for outdoor and indoor MSs in UMi environment considering two Tx power allocation schemes. The GM, which is the statistics of the \emph{average} signal-to-interference-plus-noise ratio (SINR) in the area can be given~as: \begin{align} \label{equ:geo}
{\rm GM} = \mathbb{E} \left( \frac{P_{\rm{Rx,ser}}}{N_0 + \sum_{i\neq ser}^{N_{c}}P_{\rm{Rx,i}}} \right),
\end{align} where $P_{\rm{Rx,ser}}$, and $P_{\rm{Rx,i}}$ are the received signal power from serving cell, and interference power from cell $i \ (\neq ser)$. $N_{\rm c}$ is the total number of sectors, i.e., $57$ in this evaluation. With the wrap-around architecture, MSs in the entire layout experience similar interference impact. Next, we evaluate how GM performance varies for outdoor and indoor MSs with different Tx power allocation schemes.

\subsubsection{Geometry Metric for Outdoor MSs} \label{sec:Goutdoor}

\begin{figure}
%\vspace{-3mm}
\begin{center}
\includegraphics[width=3.0in]{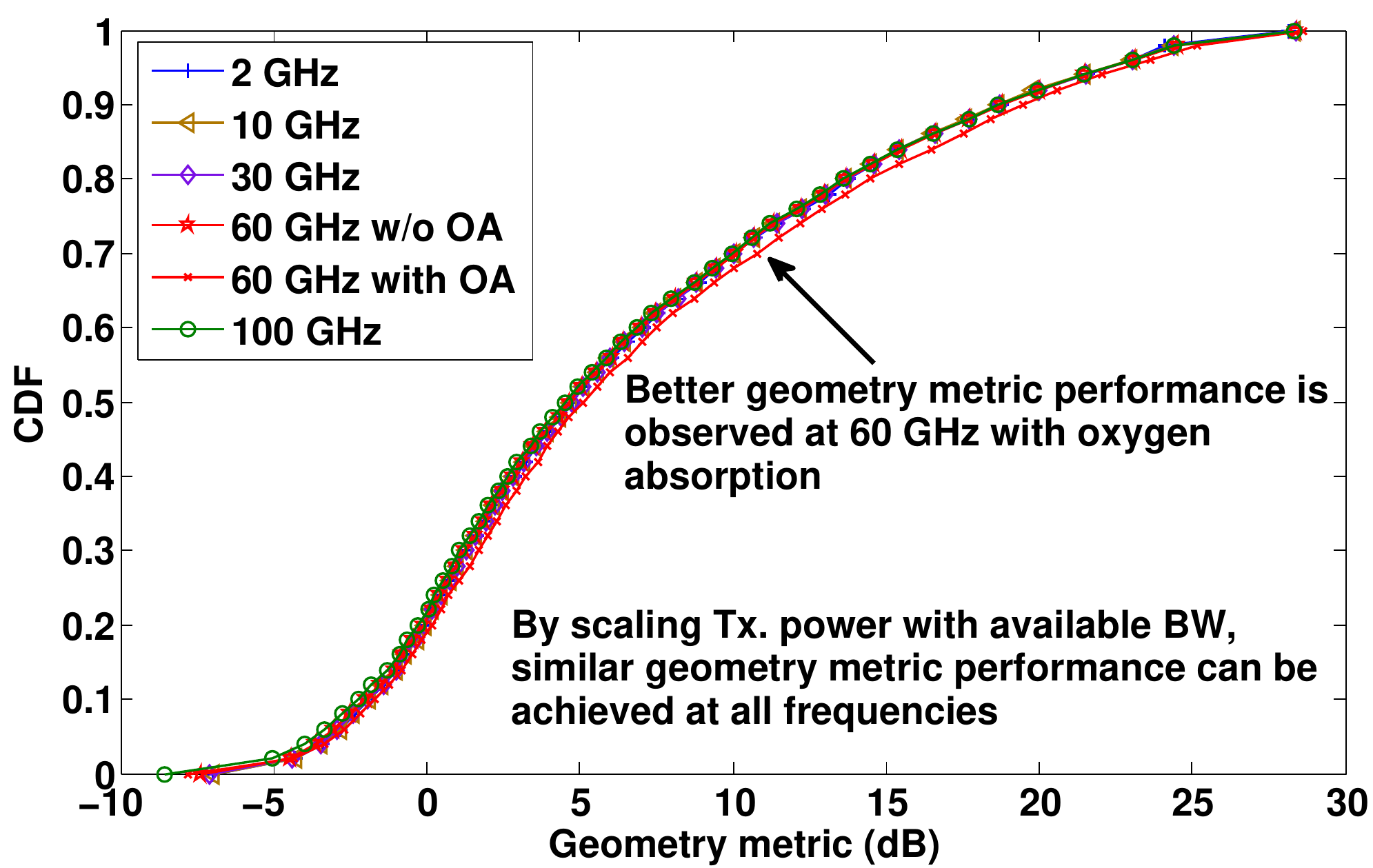}
\end{center}
%\vspace{-1mm}
\caption{CDFs of GM for outdoor MSs with scaled Tx power.}
%\vspace{-10mm}
\label{fig:OutScaleG}
\end{figure}

GM distribution for outdoor MSs with scaled Tx power (Tx power allocation scheme~1 in Table~\ref{power}), is shown in Fig.~\ref{fig:OutScaleG}. As per the figure, we can observe that GM performance is almost similar at all $f_{\rm{c}}$ and about $20 \%$ of MSs experience GM performance less than $0$~dB. The reason for experiencing similar GM at all $f_{\rm{c}}$ is due to $P_{\rm Tx}$ being scaled with available BW and as a result system tends to operate in \emph{interference-limited} condition. Further, slightly better GM performance can be observed with $L_{\rm OA}(f_{\rm{c}},d)$ at $60$~GHz compared to GM at other $f_{\rm c}$ values. This is because, decreasing of interference power due to $L_{\rm OA}$ is higher as interfering signal has to travel longer distance than that of the desired signal. Since the system is operating in \emph{interference-limited} conditions due to $P_{\rm Tx}$ being scaled with BW, reduction in interference enhances GM performance.

\begin{figure}
%\vspace{-3mm}
\begin{center}
\includegraphics[width=3.0in]{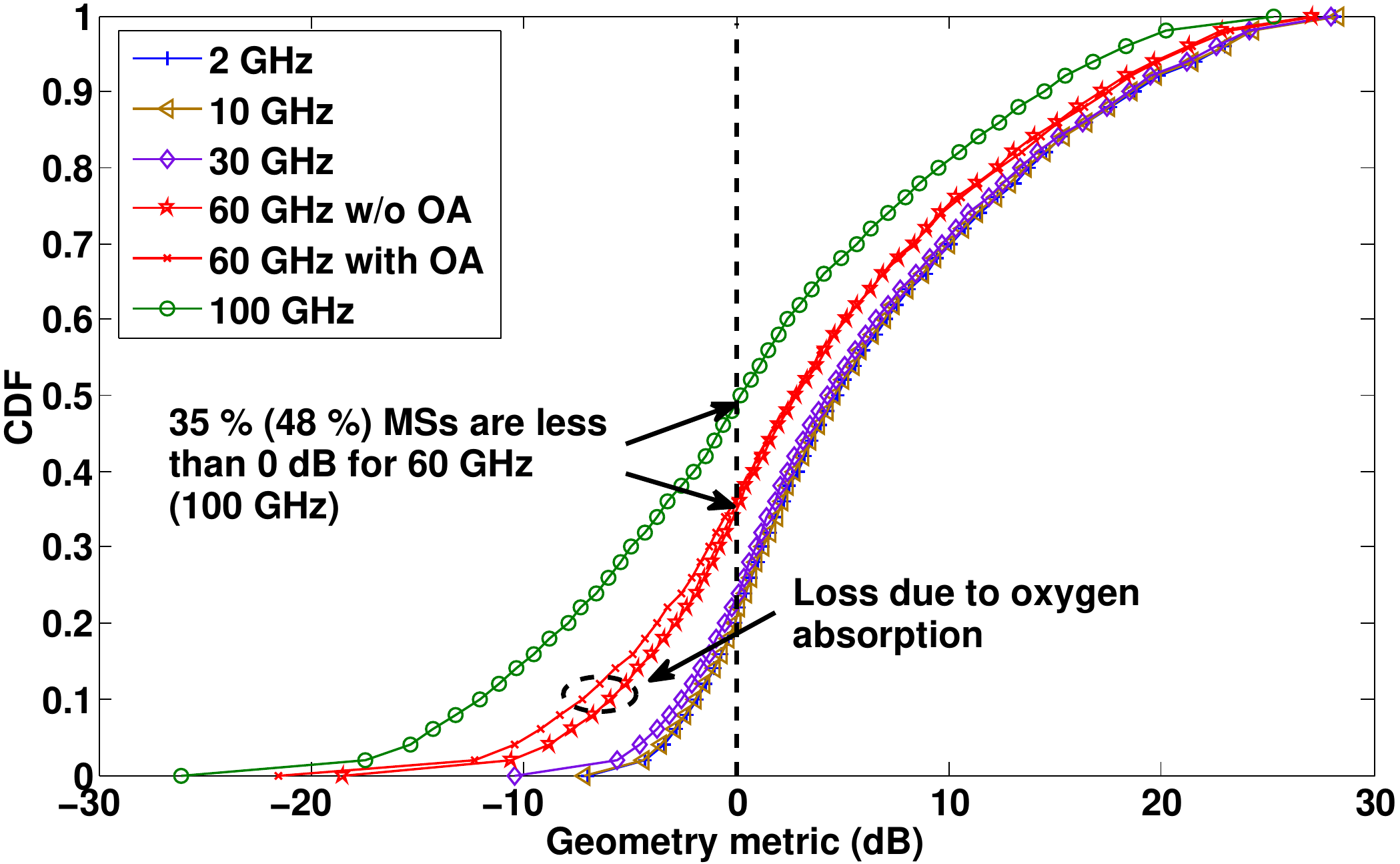}
\end{center}
%\vspace{-1mm}
\caption{CDFs of GM for outdoor MSs with constant Tx power.}
%\vspace{-4mm}
\label{fig:OutConstG}
\end{figure}

Fig.~\ref{fig:OutConstG} captures GM distribution for outdoor MSs with constant Tx power allocation. As can be observed from that, up to $30$~GHz, difference in GM performance is not very significant (about $20 \%$ of MSs experience GM less than $0$~dB), although PL is generally larger for higher frequencies. This is because, both the signal power and the interference power decrease with $f_{\rm{c}}$ without significantly impacting the GM (\emph{interference-limited}). On the other hand, for $60$~GHz, and $100$~GHz, approximately $35 \%$, and $48 \%$ MSs experience GM less than $0$~dB. The reason is, system becomes \emph{noise-limited} at these frequencies, and GM performance degrades as a result of the decreasing signal power while noise level $N_0$ stays constant.

\subsubsection{Geometry Metric for Indoor MSs}

\begin{figure}
%\vspace{-1mm}
\begin{center}
\includegraphics[width=3.0in]{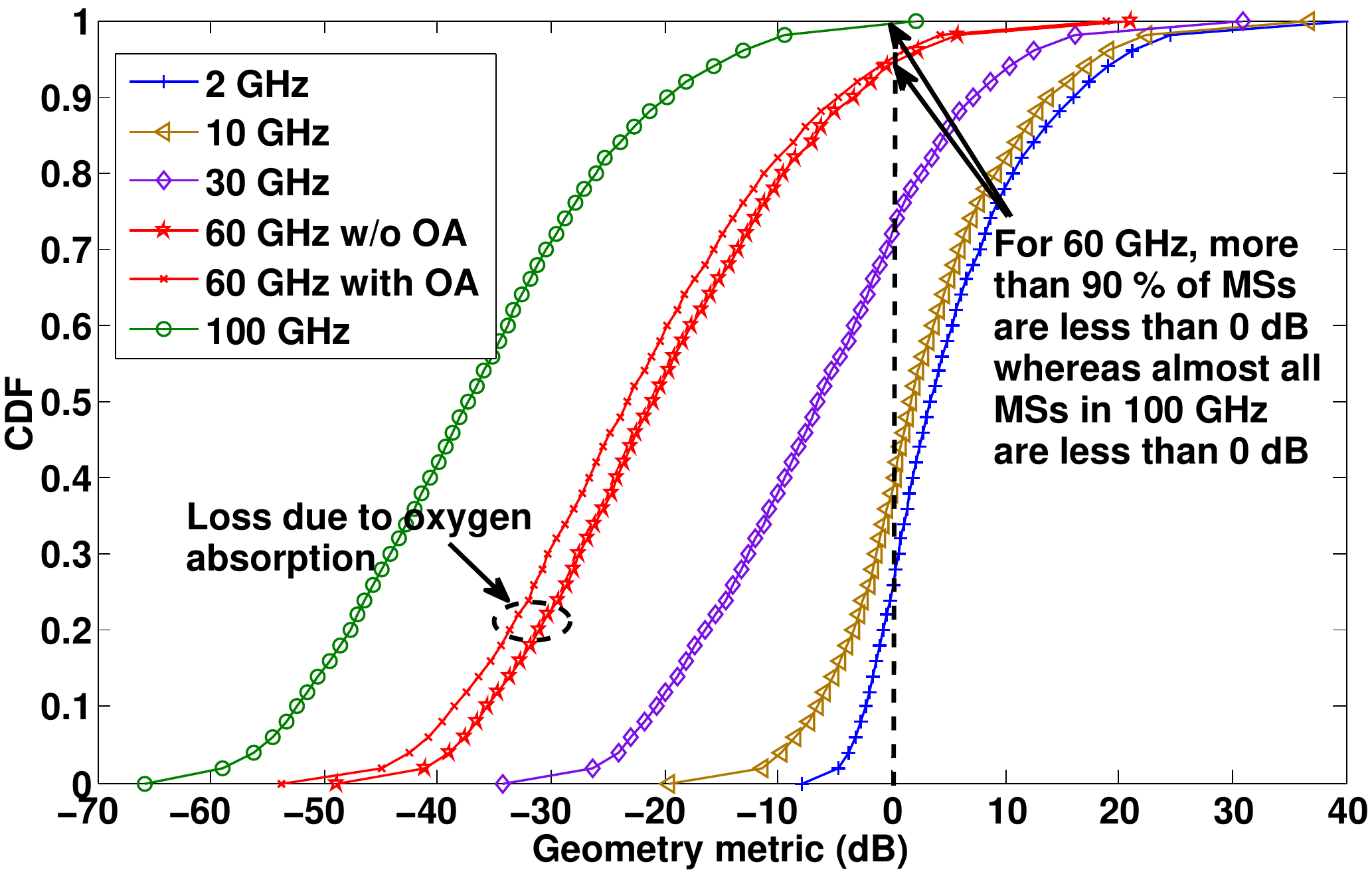}
\end{center}
%\vspace{-1mm}
\caption{CDFs of GM for indoor MSs with scaled Tx power.}
%\vspace{-1mm}
\label{fig:InScaleG}
\end{figure}

Fig.~\ref{fig:InScaleG} captures GM CDFs for indoor MSs with scaled Tx power allocation. As can be observed from that, for $30$~GHz, $60$~GHz, and $100$~GHz GM is less than $0$~dB for $75 \%$, $90 \%$ and all most all MSs, respectively. When compare this with outdoor scaled Tx power case, a clear degradation in GM performance can be observed. This is because of the larger penetration loss, $L_{\rm O2I}(f_{\rm{c}})$ experienced by indoor MSs at mmWave frequencies. Because of this $L_{\rm O2I}(f_{\rm{c}})$, as discussed in Section~\ref{inCL}, system tends to operate in \emph{noise-limited} condition at mmWave frequencies. Hence, it can be inferred that, even by scaling Tx power based on available BW, it is difficult to compensate the larger outdoor-to-indoor penetration loss incurred at higher frequencies for indoor MSs. It is generally expected that, larger PL at mmWave frequencies can be compensated partially with higher beamforming gain (with larger antenna array). However, this beamforming gain may still not be enough to overcome this outdoor-to-indoor penetration loss. Also, unlike in outdoor case, now oxygen absorption loss at $60$~GHz contributes for GM degradation as the system is in \emph{noise-limited} condition.
The GM performance is further degraded for constant Tx power allocation, even though the performance is not presented in the paper.

\section{Concluding Remarks} \label{sec:Conclusion}
In this paper, we quantitatively analyze system-level propagation performance of mmWave cellular networks using 3GPP based system-level simulation assumptions. For different
mmWave frequency bands, we study how the propagation coupling loss and geometry metric performance vary for outdoor and indoor MSs in UMi environments. Based on this investigation, it could be understood that for outdoor MSs, mmWave frequencies up to $30$~GHz are feasible options for 5G systems operation, irrespective of the Tx power allocation scheme considered. For indoor MSs,
achievable geometry metric performance at mmWave frequencies is not good enough since the system tends to operate under \emph{noise-limited} conditions, mainly due to high frequency dependent outdoor-to-indoor penetration loss.

%Compared to outdoor MSs, coupling loss performance is degraded for indoor MSs due to frequency dependent penetration loss. By scaling the Tx power based on the available BW, geometry performance degradation can be minimized for outdoor MSs at all mmWave frequencies considered. With constant Tx power allocation, up to $30$~GHz, similar geometry performance ($20 \%$ MSs less than $0$~dB) as that of scaled Tx power allocation can be observed for outdoor MSs. Hence, it can be understood that, for outdoor MSs, mmWave frequencies up to $30$~GHz are feasible options for 5G systems, irrespective of the Tx power allocation scheme considered. For indoor MSs, achievable geometry performance at mmWave frequencies even with scaled Tx power allocation, is not considerable mainly due to high frequency dependent outdoor-to-indoor penetration loss.
%\def\bibfont{\footnotesize}

%\bibliographystyle{IEEEtran}
%\bibliography{Doc_Ref}

\end{document}